\documentclass[aip, amsmath,amssymb,
 reprint,%
floatfix]{revtex4-1}

\usepackage{graphicx}
\usepackage{dcolumn}
\usepackage{bm}
\usepackage{amsmath}

\usepackage[utf8]{inputenc}
\usepackage[T1]{fontenc}
\usepackage{mathptmx}
\usepackage{etoolbox}
\usepackage{color,soul}

{}

\makeatletter
\def\@email#1#2{%
 \endgroup
 \patchcmd{\titleblock@produce}
  {\frontmatter@RRAPformat}
  {\frontmatter@RRAPformat{\produce@RRAP{*#1\href{mailto:#2}{#2}}}\frontmatter@RRAPformat}
  {}{}
}%
\makeatother
\begin{document}

\preprint{AIP/123-QED}

\title[Ultrafast coherent excitation of an Ytterbium ion with single laser pulses]{Ultrafast coherent excitation of an Ytterbium ion with single laser pulses}
\author{Kenji Shimizu}
\affiliation{Centre for Quantum Dynamics, Griffith University, Brisbane, 4111 QLD, Australia}

\author{Jordan Scarabel}%
\affiliation{Centre for Quantum Dynamics, Griffith University, Brisbane, 4111 QLD, Australia}

\author{Elizabeth Bridge}
\affiliation{Australian Research Council Centre of Excellence for Engineered Quantum Systems, School of Mathematics and Physics, University of Queensland, St Lucia, Queensland 4072, Australia}

\author{Steven Connell}%
\affiliation{Centre for Quantum Dynamics, Griffith University, Brisbane, 4111 QLD, Australia}

\author{Moji Ghadimi}%
\affiliation{Centre for Quantum Dynamics, Griffith University, Brisbane, 4111 QLD, Australia}

\author{Ben Haylock}
\affiliation{Centre for Quantum Dynamics, Griffith University, Brisbane, 4111 QLD, Australia}

\author{Mahmood Irtiza Hussain}%
\affiliation{Institute for Quantum Optics and Quantum Information (IQOQI), Austrian Academy of Sciences, Innsbruck 6020, Austria}
\affiliation{Institute for Experimental Physics, University of Innsbruck, Innsbruck 6020, Austria}

\author{Erik Streed}%
\affiliation{Centre for Quantum Dynamics, Griffith University, Brisbane, 4111 QLD, Australia}
\affiliation{Institute for Glycomics, Griffith University, Gold Coast, QLD 4222, Australia}

\author{Mirko Lobino}%
\affiliation{Centre for Quantum Dynamics, Griffith University, Brisbane, 4111 QLD, Australia}
\affiliation{Queensland Micro and Nanotechnology Centre, Griffith University, Brisbane, QLD 4111, Australia}
\email{m.lobino@griffith.edu.au}

\date{\today}

\begin{abstract}
Experimental realizations of two qubit entangling gates with trapped ions typically rely on addressing spectroscopically-resolved motional sidebands, limiting gate speed to the secular frequency. Fast entangling gates using ultrafast pulsed lasers overcome this speed limit. This approach is based on state-dependent photon recoil kicks from a sequence of counter-propagating, resonant, ultrafast pulse pairs, which can allow sub-microsecond gate speeds. 
Here we demonstrate a key component of the ultrafast gate protocol, the coherent excitation of a $^{171}$Yb$^+$ ion across the $^2$S$_{1/2}$-$^2$P$_{1/2}$ transition with a single near-resonant short optical pulse at 369.53~nm. We achieve a maximum population transfer of 94.3(6)\% using a picosecond pulsed laser that can be tuned across the $^2$S$_{1/2}$-$^2$P$_{1/2}$ transition, and 42.53(13)\% with 190(7)~GHz detuning.   
\end{abstract}

\maketitle


Trapped-ion qubits are a leading candidate for the realization of a full-scale quantum computer because of their long coherence time, deterministic gates and readout, and their potential for scalability\cite{Bruzewicz2019,Bermudez2017}. High fidelity two-qubit gates have been demonstrated on this platform\cite{Schmidt2003,Gaebler2016,Ballance2016}, with a speed on the order of few microseconds. This is because most two-qubit gates\cite{Cirac1995, Molmer99} are performed in the Lamb-Dicke regime via the excitation of single motional sidebands, tying the gate duration to the period of the secular motion of the ions, typically on the order of microseconds. These gate protocols\cite{Cirac1995, Molmer99} require near ground state motional cooling and for some also individual laser addressability for each ion\cite{Cirac1995}. Furthermore the speed limit set by secular motion worsens for ion crystals since the frequency of the collective motional modes decreases with the number of ions and cross talk between modes can reduce the gate fidelity\cite{Monroe2013}.

In 2003, García-Ripoll et al.\cite{Garcia2003} proposed a scheme for a sub-microsecond two-qubit gate that does not require individual ion-laser addressability or ground state motional cooling. The scheme operates by driving the ions' optical transition with counter-propagating, resonant, ultrafast $\pi$-pulses that excite and then de-excite the ion. The pulse pairs are polarized to deterministically result in either a momentum kick of 2$\hbar$k or no momentum kick, depending on the internal qubit state of the ion. The sequence of 2$\hbar$k kicks generates closed state-dependent trajectories of the ions through their phase-space where they all go back to their initial motional and qubit states at the end of the protocol\cite{Bentley2015}. Pairs of qubits undergo a phase shift from the state-dependent kicks, only if they are both in a particular state, resulting in a two-qubit phase gate\cite{Bentley2015,Bentley2013}.

Other protocols have been used to realize single ion gates at a rate faster than the secular motion\cite{Campbell2010}, and one-qubit spin-motion entanglement\cite{Mizrahi2013}. One of these two-qubit gate protocols\cite{Schafer2018} uses amplitude-shaped Raman pulses to drive the motion of the ions along trajectories in phase-space so that the gate operation becomes insensitive to the optical phase of the pulses. Although this approach achieved a gate time of 1.6~$\mu$s with a fidelity of 99.8\%, it only works in the Lamb-Dicke regime. Another gate protocol\cite{Wong2017} demonstrated the generation of the entangled Bell state with a fidelity of 76\% using multiple Raman pulse trains to create a single spin-dependent kick. While this method doesn’t require confinement of ions to within the Lamb-Dicke regime, three optical delay stages split single pulses into eight sub-pulses which compound in infidelity for a single state-dependent force, leading to the low fidelity of the entangling gate. 

Implementation of the fast gate protocol\cite{Garcia2003,Bentley2015,Bentley2013} using resonant pulses is at the stage where suitable ultrafast laser sources have been realized for calcium\cite{Hussain2021} and ytterbium\cite{Hussain2016} ions, and demonstration of coherent excitation of $^{40}$Ca$^+$ ions has being reported\cite{Heinrich2019}. Here we show the coherent excitation of a $^{171}$Yb$^+$ ion along the $^2$S$_{1/2}\rightarrow^2$P$_{1/2}$ transition using a single resonant $\pi$-pulse. Using picosecond pulses, which are much shorter than the 8.12(2)~ns lifetime\cite{Olmschenk2009} of the excited state, we achieve a population inversion of 94.3(6)\% when the laser is near resonance and 42.53(13)\% with 190(7)~GHz detuning. We compared our results with the semi-classical two-level atom theory and found good agreement with our temporal and spectral laser parameters.

\begin{figure*}
\includegraphics[width=\linewidth]{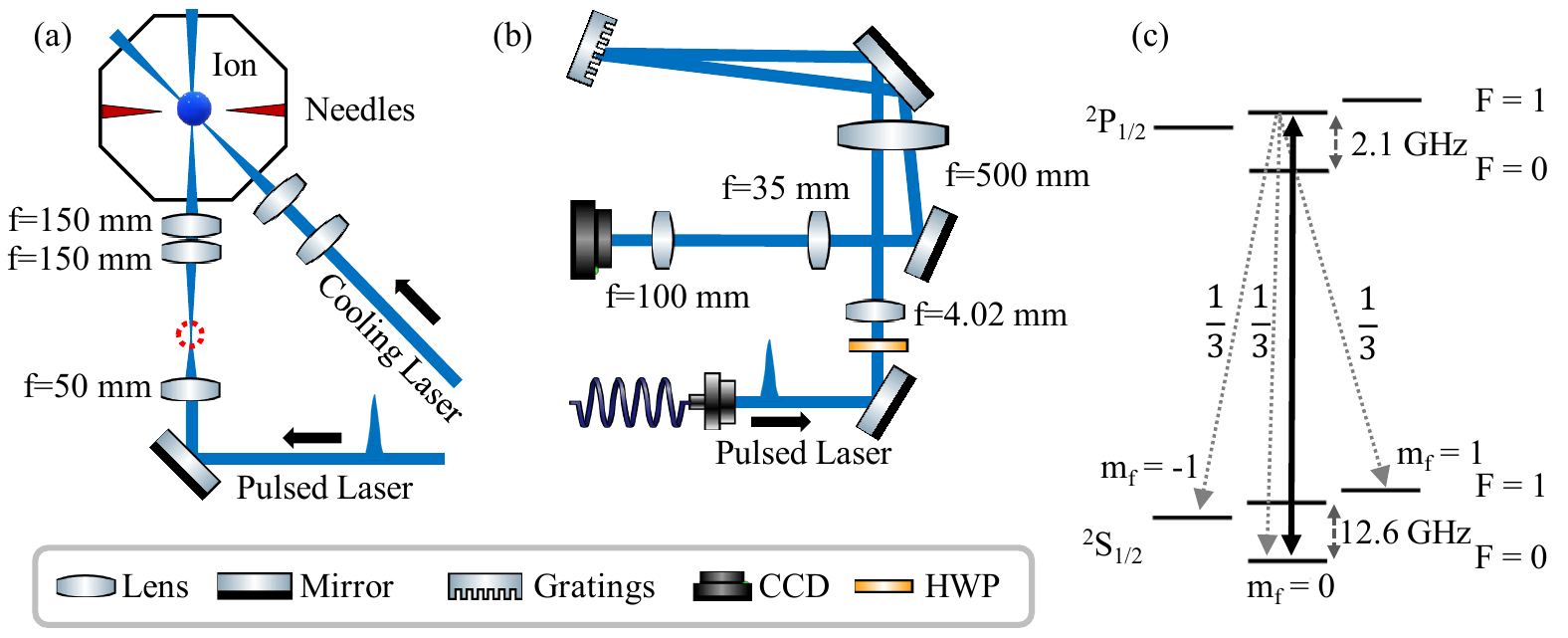}
\caption{\label{fig:schem}  
(a) The pulsed laser and cooling laser are focused onto an ion in an vacuum chamber using the illustrated optics. Due to geometric limitations we use an f-2f lens configuration where a focused spot outside the chamber (indicated with a red dotted circle) can be replicated at the ion inside the chamber with a relay of two f=150 mm lenses. (b) The pulsed laser was spectrally analysed with a diffraction grating spectrometer setup. The light coming from a fiber is collimated with a f=4.02~mm  aspheric lens, expanded by a telescope with magnification M=124, and sent to a holographic grating. The reflected pulses are slightly offset from the input path so they can be picked off using a D-shaped mirror and measured by a CCD camera. (c) The $^2$S$_{1/2}$-$^2$P$_{1/2}$ energy level diagram shows blue arrows illustrating the transition the pulsed laser is resonant with. The pulsed laser has linear polarization parallel to an external magnetic field, so it can excite ions prepared in the $^2$S$_{1/2}~F=0$ state only to the $^2$P$_{1/2}~F=1, m_F=0$ state via a $\pi$-transition. Dotted arrows show the  spontaneous emission paths.}
\end{figure*}

A single $^{171}$Yb$^+$ ytterbium ion is trapped and Doppler cooled in a quadrupole Paul trap formed by two tungsten needles driven by radio frequency high voltage at 18.810~MHz (see Fig.~\ref{fig:schem}a). The ion is cooled by a CW laser, which is red-detuned by 10~MHz from the $^2$S$_{1/2}$ F=1 $\leftrightarrow$ $^2$P$_{1/2}$ F=0 transition\cite{Ghadimi2020}, and repumped from the $^2$S$_{1/2}$ F=0 dark state with light from 14.7 GHz sidebands generated by a Qubig PM-Yb171+\_14.7M2 electro-optic modulator (EOM). The ion's fluorescence is collected by a 0.64 numerical aperture (NA) binary phase Fresnel lens, located on top of the ion but inside vacuum chamber, and measured with a photo-multiplier tube (PMT) mounted above the chamber. An additional 935 nm laser with 3.07 GHz sidebands generated by diode current modulation repumps the atom out of the meta-stable $^2$D$_{3/2}$ state. Details of the trap geometry can be found in~\textcite{Blums18}.

The laser used for coherent excitation of the ion is a home-built system\cite{Hussain2016} that uses a telecom band femtosecond fiber laser as the master oscillator. The oscillator wavelength is centred at 1564~nm, and the repetition rate is frequency stabilized at 300.00000~MHz. Femtosecond pulses are used to generate supercontinuum in a highly nonlinear fiber, where a portion around 1108.6~nm is selected with three chirped fiber Bragg grating (CFBG) mirror stages and amplified. Second harmonic generation (SHG) in a periodically poled stoichiometric lithium tantalate (PPSLT) crystal produces radiation around 554 nm. Sum frequency generation between the SHG and 1108.6~nm beam in a LiB$_3$O$_5$ (LBO) crystal produces laser pulses around $\lambda$=369.52~nm. A maximum average power of 100~mW in the UV is generated which can be frequency tuned around the $^2$S$_{1/2}\leftrightarrow^2$P$_{1/2}$ transition (Fig.~\ref{fig:schem}c) by adjusting the temperatures of the CFBGs and the PPSLT crystal independently. The final pulse is near transform limited with a duration of $\sim$1~ps. 

During experiment, the UV pulse spectrum is monitored by a high resolution spectrometer\cite{Connell2021} built around a custom Newport Richardson diffraction grating mirror with 4320~grooves/mm and dimensions of 128~mm$\times$102~mm (see Fig.~\ref{fig:schem}b). The optically magnified pulsed laser illuminates the diffraction grating mirror in Littrow configuration, where the laser is diffracted on a slightly different trajectory and then sent to a CCD camera. The spectrometer has a resolution of 3.6(2) GHz around 370~nm and was calibrated against the cooling laser. For the ion excitation, a collimated beam of 1.875~mm waist and a beam quality factor M$^2$ of 2.7, is focused down to a waist of 8.5~$\mu$m with a lens of 50~mm focal. We relay that spot size on the ion with a 75~mm focal length relay lens set-up (Fig.~\ref{fig:schem}a).

\begin{figure}[t]
\includegraphics{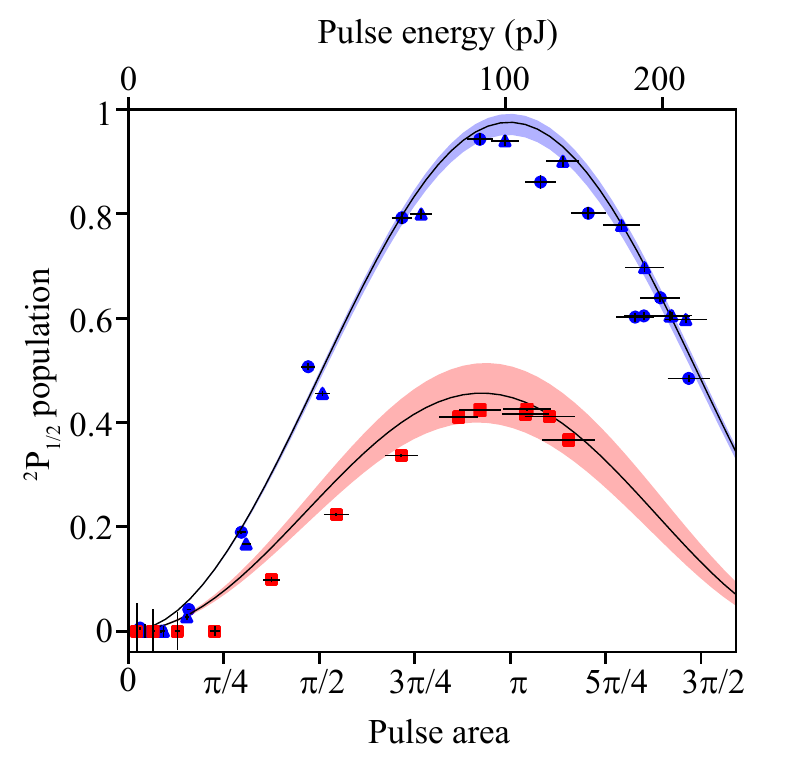}
\caption{\label{fig:flop} Excitation probability of the ion to $^2$P$_{1/2}$ with single pulses as a function of pulse area $\Theta=\Omega_{\mbox{eff}}\times t_{\mbox{eff}}$ (lower axis) and the pulse energy (upper axis). Blue data points represent on-resonant excitation measurements, where triangles and then circles are two consecutively taken sets of data. Red data points are off-resonant excitation measurements also taken with the same ion. The blue curve correspond to a laser detuning of 33(7)~GHz and $t_{\mbox{eff}}=2.36(1)$~ps, while the one in red is for 190(7)~ GHz detuning and $t_{\mbox{eff}}=2.25(2)~$ps. The error bars on the experimental data represent two standard deviations for horizontal and vertical axes, where the main source of horizontal error is from fluctuations in pulse intensity on the ion. The vertical error bar is from population parameter estimation in fitting measurements' statistical distributions. The theoretical curves have confidence bands of two standard deviations. The main sources of error are from pulse duration and laser detuning.}
\end{figure}

Coherent excitation measurements are performed with a four step protocol of 289~$\mu$s duration repeated every 426.66~$\mu$s. First the ion is Doppler cooled for 40~$\mu$s and then optically pumped to the $^2$S$_{1/2}~F=0$  state for 20~$\mu$s by applying 2.1 GHz sidebands to the cooling beam generated by a QUBIG PM-Yb171+\_2.1G EOM. The ion is then excited with a single laser pulse and left to decay to the ground state for 2~$\mu$s (Fig.~\ref{fig:schem}c). This is much longer than the 8.12(2)~ns lifetime of the excited state\cite{Olmschenk2009}. After excitation, the ion can decay into the states $^2$S$_{1/2}~F=1~m_F=\pm1$ and $^2$S$_{1/2}~F=0$, with a probability of 1/3 each. Read-out via state selective fluorescence\cite{Ejtemaee2010} is performed with the Doppler beam for 227~$\mu$s\cite{Wolk2015}. If the ion was excited by the single pulse and had spontaneously decayed, the ion's atomic state will be in the $^2$S$_{1/2}~F=1$ manifold with a probability of 2/3 (Fig.~\ref{fig:schem}c) and be bright during state read-out. 
We measure the probability of single pulse excitation as a function of the pulse energy. For each value of pulse energy, we repeat the protocol approximately 68500 times to build up statistics for the probability of excitation into the $^2$P$_{1/2}$ state. The count statistics are fitted to two complimentary weighted probability mass functions \cite{Wolk2015} that describe the bright and dark count distributions. The weight of the bright component is our estimate of the $^2$S$_{1/2}~F=1$ manifold population. The population measurement of the $^2$S$_{1/2}~F=1$ manifold is scaled by 3/2 to account for the $^2$P$_{1/2}~F=1$ branching ratio. The results are shown in Fig.~\ref{fig:flop}, including the measurements performed with a laser detuning of 190(7)~GHz. The maximum population transfer is 94.3(6)\% for on-resonance case and 42.53(13)\% with detuning.

\begin{figure}[htbp]
\includegraphics[width=\linewidth]{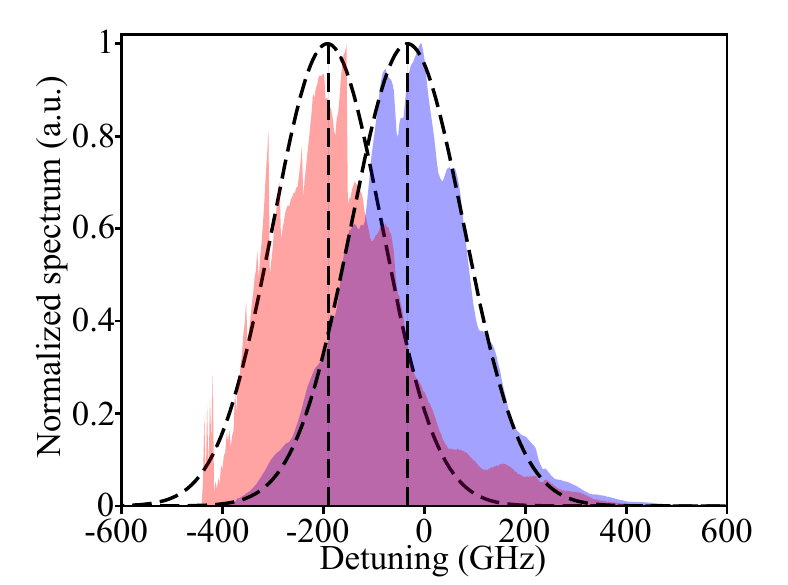}
\caption{\label{fig:spec} Averaged frequency spectrum on-resonance (blue) and off-resonance (red). The black dashed lines show the Gaussian fit of the spectra and their centre frequencies.}
\end{figure}

During the acquisition, which took a total of 65~min, the laser spectrum was measured every 10~s from the high resolution spectrometer (see Fig.~\ref{fig:schem}b). Figure~\ref{fig:spec} shows the average of the measured spectra for the on-resonance and detuned case. Both curves are fitted with a Gaussian function from which we find detunings -33(7)~GHz and -190(7) GHz from the $^2$S$_{1/2} F=1\rightarrow^2$P$_{1/2}~F=0$ transition for the on-resonant and detuned excitation respectively.

The experimental data shown in Fig.~\ref{fig:flop} are fitted with the equation for rectangular pulse excitation
\begin{equation}\label{eq:pop}
P_{ex}=\frac{\Omega_{\mbox{eff}}^2}{\Omega_{\mbox{eff}}^2+\Delta^2}\sin^2\frac{1}{2}t_{\mbox{eff}}\sqrt{\Omega_{\mbox{eff}}^2+\Delta^2}
\end{equation}
where $\Delta$ is the detuning measured from spectrometer, $t_{\mbox{eff}}$ the effective pulse duration, and $\Omega_{\mbox{eff}}$ the effective Rabi frequency. The theoretical curves in Fig.~\ref{fig:flop} were obtained by defining the  $\Omega_{\mbox{eff}}=\alpha\sqrt{C_{sc}}$, with $C_{sc}$ the number of background scattering counts from the single pulses, which is directly proportional to the pulse energy, and $\alpha$ the only fitting parameter. The pulse duration was calculated from the spectra of Fig.~\ref{fig:spec}, giving a $1/e^2$ field duration of
0.941(4)~ps for the on-resonance and 0.896(9)~ps for the detuned case. These pulse durations are in good agreement with the 0.88(5)~ps measured independently with a Michelson interferometer\cite{Levchenko2006}. These values are multiplied by $\sqrt{2\pi}$ in order to get the equivalent square pulse area, corresponding to an effective pulse duration $t_{\mbox{eff}}=2.36(1)$~ps for the on-resonance case and $t_{\mbox{eff}}=2.25(2)$~ps for the off-resonance one. 

From the on-resonance fitting curve shown in Fig.~\ref{fig:flop} we estimate a Rabi frequency for the maximum population transfer $\Omega_{\mbox{eff}}=1.226(11)\times10^{12}$~rad/s. This value is compared with a theoretical estimation obtained from
\begin{equation}
\label{eq:rabi}
\Omega_{\mbox{th}} = C \Gamma \sqrt{\frac{I_p}{2  I_{s}}}
\end{equation}
where $\Gamma=2\pi\times19.6$~MHz is the natural linewidth of the $^2$S$_{1/2}$ - $^2$P$_{1/2}$ transition in $^{171}$Yb$^+$, $C=1/\sqrt{3}$ the Clebsch–Gordan coefficient, $I_{s}=508$~W/m$^2$ the saturation intensity, and $I_p$ the peak intensity of the pulsed laser. For a consistent comparison we assume that our pulses are square with a duration $t_p=\sqrt{\pi}\times$ 0.941(4)~ps=1.667(7)~ps, and peak intensity $I_p$ given by 
\begin{equation}
I_p = \frac{2 E}{\pi w_0^2  t_p}=458(9)\times10^{9}\mbox{ W/m}^2
\end{equation}
where, $E$=0.0867(17)~nJ is the measured energy of a pulse, and w$_0$=8.5~$\mu$m the beam waist at the ion location. With these numbers we estimate a theoretical Rabi frequency $\Omega_{\mbox{th}}=1.510(15)\times10^{12}$~rad/s, which is in good agreement with the value obtained from our measurements.



In conclusion, we have experimentally demonstrated ultrafast coherent excitation of $^{171}$Yb$^+$ in an ion trap with single laser pulses. The maximum population transfer we achieve in this work is 94.3(6)\% and can be improved by having finer control over the detuning of the pulsed laser relative to the atomic resonance. We also have developed a technique to monitor the pulse spectrum using a high resolution diffraction grating spectrometer. This result will pave the way towards demonstrating counter-propagating $\pi$-pulse schemes, which is a fundamental requirement for the demonstration of a fast entangling gate.

\begin{acknowledgments}
We thank Dr Valdis Bl\=ums for helpful discussions and comments. This research was financially supported by the Griffith University Research Infrastructure Program, the Griffith Sciences equipment scheme, Australian Research Council Linkage project (LP180100096M), ML was supported by the Australian Research Council Future Fellowship (FT180100055), JS, SC, and KS were supported by the Australian Government Research Training Program Scholarship. 
\end{acknowledgments}

\section*{Data Availability}
The data are available from the corresponding author on reasonable request.

\section*{References}

\bibliography{Shimizu_ref}

\end{document}